\documentstyle[12pt,emlines2]{article}
\begin{document}

\voffset=-39mm
\hoffset=-16mm
\def\a{\alpha }   \def\b{\beta }
\def\g{\gamma }   \def\G{\Gamma }
\def\t{\theta }                    \def\S{\Sigma}
\def\d{\delta }   \def\D{\Delta }  \def\e{\epsilon }
\def\o{\omega }   \def\O{\Omega }
\def\l{\lambda }     \def\L{\Lambda }    \def\k{\kappa}
\def\ni{\noindent }
\def\und{\underline}

\font\Cal=msbm10 scaled \magstep1
\font\Calscr=msbm7 scaled \magstep1
\font\Calscrscr=msbm5 scaled \magstep1
\newfam\Calfam
\textfont\Calfam=\Cal
\scriptfont\Calfam=\Calscr
\scriptscriptfont\Calfam=\Calscrscr
\def\Cal{\fam\Calfam}

\rightline  {hep-ph/9712411 \hspace{6mm}}
\bigskip
\medskip

\begin{center}
\Large{\bf Quantum Groups in Hadron Phenomenology}
\end{center}

\medskip
\bigskip
\begin{center}
{\bf A.M.~Gavrilik}
\bigskip \\
{\it Bogolyubov Institute for Theoretical Physics,  \\
     Metrologichna str. 14b, Kiev-143, Ukraine}
\end{center}

\bigskip
\bigskip
\begin{abstract}
We show that application of quantum unitary groups,
in place of ordinary flavor $SU(n_f),$ to such static aspects
of hadron phenomenology as hadron masses and mass formulas
is indeed fruitful. The so-called $q$-deformed mass formulas are given
for octet baryons $\frac12^+$ and decuplet baryons $\frac32^+,$
as well as for the case of vector mesons $1^-$ involving heavy flavors.
For deformation parameter $q,$ rigid fixation of values is used.
New mass sum rules of remarkable accuracy are presented.
As shown in decuplet case, the approach accounts for
effects highly nonlinear in $SU(3)$-breaking.
Topological implication (possible connection with knots) for
singlet vector mesons and the relation $q\leftrightarrow \theta_{\rm C}$
(Cabibbo angle) in case of baryons are considered.
\end{abstract}
\bigskip
\medskip

\bigskip
\ni{\bf 1. Introduction}
\bigskip

During last decade, quantum groups and
quantum (or $q$-) algebras [1] have shown their apparent
effectiveness in diverse problems of theoretical physics,
see overviews [2]. In particular, one gets essential improvement
of phenomenological description of superdeformed nuclear bands and
spectra of diatomic molecules by replacing [3]
usual $su(2)$-symmetry (underlying rigid-rotator based
description of rotational spectra) with its $q$-deformed analogue
$su(2)_q.$

More recently, in the context of hadron phenomenology the use of
quantum groups/algebras has been proposed [4-8]. Here we discuss
some results and implications of such an application.
Basic idea of [4-7] consists in adopting the
$q$-{\it deformed version of flavor symmetries}
(staying formally within the first-order flavor symmetry breaking)
in order to get better agreement with empirical data for hadron
masses.    That is, we start by replacing the usual
(isospin and higher) unitary symmetries with their quantum counterparts:
$SU(n_f)\rightarrow SU_q(n_f),$ $\ n_f\ge 2.$
Main motivation of such replacement comes from the following two facts:

\smallskip
\hskip 10mm \parbox{144mm}{ $\bullet$  Successful application
         of the $q$-algebra $su(2)_q$ for phenomenological
        description of rotational bands of (super)deformed nuclei
        and diatomic molecules [3]; }

\vskip 2mm
\hskip 10mm \parbox{144mm} { $\bullet$ The fact known from
      representation theory of quantum groups and $q$-algebras
      that each finite-dimensional irreducible representation (irrep)
      of $SU(n)$ smoothly $q$-deformes [9], at $q$ not
      a root of unity, to respective
      irrep of $SU_q(n)$ of the same dimension. }

\vskip 2mm
\ni More precisely, we exploit the $q$-algebras $U_q(su_n)$
corresponding to $SU_q(n),$ along with their irreps,
as flavor symmetries of hadrons - vector mesons $1^-$ and
baryons ${1\over 2}^+,\ $ ${3\over 2}^+.$
To calculate hadron masses, within our model we utilized [4-7]
simple and natural but sufficiently effective method of performing
necessary (flavor) symmetry breaking. This method directly extends
to a $q$-deformed case the approach, based on the concept of
unitary/pseudounitary dynamical groups, earlier used in order to
treat hadron masses and mass formulas described with conventional
groups $SU(n)$ of flavor symmetries. The virtue of method is that it
allows to bypass difficulties related with $q$-CGCs and $q$-Casimirs
which for higher rank quantum groups appear rather nontrivial.

With the help of appropriate $q$-algebras $U_q(u_{n+1})$ or
$U_q(u_{n,1})$ of 'dynamical' symmetry, one realizes necessary
breaking of $n$-flavor symmetries up to exact (for strong
interactions alone) isospin symmetry $su_q(2)_I$ and obtains
the $q$-analogues of mass relations (MRs) [4-7].
From these $q$-analogues of hadron mass formulas, in
the non-deformed (`classical') limit $q\to 1,$ one recovers
the familiar hadronic (Gell-Mann--Okubo, or GMO, and equal spacing)
mass sum rules [10]. In this point our approach principally differs from
that of Ref.[8], wherein classical limit implies complete degeneracy
of masses both within octet and within decuplet.

At definite values of deformation parameter $q,\ q\ne 1,$ the
$q$-deformed baryon mass formulas produce new mass sum rules,
both octet and decuplet ones, which hold
with better accuracy than classical GMO sum rule
and equal-spacing rule respectively.

 In the case of vector mesons it turns out that
all the $q$-dependence in expressions for masses and in
resulting $q$-deformed MRs is expressible [4,7] in terms of certain
Lorant-type polynomials (of $q$) related with invariants
of respective torus knots. As a consequence, concrete torus
knots can be associated with concrete vector quarkonia.
Moreover, a possibility appears to {\it distinguish different
flavors}, through their vector quarkonia masses,
{\it by topological means} - that is, using
'braid overcrossing number' or 'torus winding number',
see section 3.

Already in [4] it was clear that deformation parameter $q$
'regulates' the issue of singlet meson mixing. Say, instead of
manifest introducing the $\phi$--$\o$ mixingle angle, usual for
ordinary $SU(3)$-based approach, one gets agreement with data
of meson mass sum rule (MSR) which involves the {\em physical}
$\phi$-meson, not mixed state, merely due to adequate value of $q$.
This correlation: deformation parameter $\leftrightarrow$ mixing angle
goes even further in the case of fermions (baryons).

In the framework of fundamental problem of fermion masses and mixings
it is well-known that the Cabibbo angle $\t_{\rm C}$
can be directly related with fermion (quark) masses: the original
formula [11] $\tan^2\theta_{\rm C} = {m_d}/{m_s}$
was subsequently reobtained within different approaches including
thoses based on supersymmetry or noncommutative geometry.
As will be demonstrated in sections 4-6, the two values of deformation
parameter whose fixation in the $q$-deformed mass formulas for
octet and decuplet baryons provides us with two new
mass sum rules of remarkable precision, are connectible
in a simplest possible way with the Cabibbo angle.

\bigskip
\ni {\bf 2. $q$-Deformed mass formulas for vector mesons}
\bigskip

Details of derivation are contained in [4,7]. Here we only recall
the basic points (say, for 3 flavors): (i) Assign vector meson states
from octet (isotriplet, two isodoublets, iso\-singlet) with the
corresponding vectors of orthonormal Gelfand-Tsetlin basis, e.g.,
$ \vert \rho \rangle = \vert {\{ 8\} }_3; {\{ 3\} }_2;{\alpha }_{\rho }
\rangle, \ \  \vert {\omega }_8\rangle =\vert {\{ 8\} }_3;{\{ 1\} }_2;
{\alpha }_8\rangle, $ where $\alpha_{\rho}$ labels the charge
states within isotriplet; (ii) Embed the octet of $U_q(u_3)$
into adjoint 15-plet representation of dynamical $U_q(u_4)$;
(iii) Take mass operator for $U_q(u_3)$ symmetry breaking in terms
of appropriate generators of $U_q(u_4)$, namely
$\hat {M_3} = M_0 +{\g }_3(A_{34}A_{43} + 
                                A_{43}A_{34});$
(iv) Calculate matrix elements
$\langle\rho\vert \hat {M_3}\vert\rho\rangle ,
\langle {\omega}_8\vert\hat {M_3}\vert {\omega}_8\rangle$, etc. .

In more general case, $3\le n\le 6,\ $ we use Gelfand-Tsetlin basis
for meson states from $(n^2-1)$-plet of 'flavor' $U_q(u_n)$ embedded
into $\{(n+1)^2-1\}$-plet of 'dynamical' $U_q(u_{n+1});\ $
mass operator, commuting with the 'isospin and hypercharge'
$q$-algebra $U_q(u_2)$, is constructed (bilinearly) from
relevant generators of 'dynamical' algebra $U_q(u_{n+1}),\ $
and has the form [4,7] which agrees with the concept of
symmetry breaking due to quark mass differences.

The expressions for masses, obtained from calculations,
depend on the symmetry breaking parameters like
$\g_3 $ and on the deformation parameter $q$.
For example,
\vskip -1mm
$$
m_{\rho}=M_o,  \hskip 12mm  m_{K^*}=M_o -{\gamma}_3,  \hskip 12mm
m_{{\omega}_8}= M_o - 2\ {[2]_q \over [3]_q}\ {\gamma}_3,    \eqno{(1)}
$$
\vskip -1mm
$$m_{D^*}=M_o +{\gamma}_4,     \hskip 22mm
  m_{F^*}=M_o -{\gamma}_3 +{\gamma}_4,   \hskip 20mm
$$
$$m_{{\omega}_{15}}= M_o + 2 \biggl ({4\over [2]_q}-{[3]_q\over [4]_q}
-{[4]_q\over [3]_q}\biggr ) {\gamma}_3 + 2\
{[3]_q\over [4]_q}\ {\gamma}_4,        \hskip 20mm      \eqno{(2)}
$$
\vskip 1mm
\ni where $[x]_q \equiv [x]\equiv (q^x - q^{-x})/(q - q^{-1})$
is the $q$-number $[x]$ corresponding to usual number $x,$ and
the requirement that (isodoublet) particles and their anti's
have equal masses was taken in account. An important fact
is that $q$-{\it dependence appears only in masses of}
${\omega }_8$, ${\omega }_{15}$,
${\omega }_{24}$, ${\omega }_{35}$
(isosinglet within octet and singlets within
$(n^2-1)$-plets of $U_q(u_n),\ n=4,5,6 $).

Exclusion of unknown parameters results in the
$q$-deformed mass relations [4,7]
$$
[n]_{(q)}\  m_{\omega_{n^2-1}} + (b_{n;q}+2n-4) \ m_{\rho} =
2\ m_{D_n^*} + (c_{n;q} + 2) \sum_{r=3}^{n-1}m_{D_r^*}  \eqno (3)
$$
where the denotion $[n]_q/[n-1]_q \equiv [n]_{(q)}$ is used and
$$
b_{n;q}\equiv n \ c_{n;q} -6\ [n]^2_{(q)} +
\Bigl ({24\over [2]_q}-1\Bigr )[n]_{(q)}        \hskip 12mm
 c_{n;q}\equiv 2 \ [n]^2_{(q)}-{8\over [2]_q}[n]_{(q)}.
$$
These $q$-analogues show that
coefficients at masses are obtained from their 'classical'
prototypes in a more complex way than merely by replacing
$a \rightarrow [a]_q$.

At $n=3$ the Eq.(3) contains the $q$-analogue of GMO relation
$$
m_{{\omega }_8} + \Bigl (2 {{[2]_q}\over {[3]_q}}
- 1\Bigr ) m_{\rho } = 2 {{[2]_q}\over {[3]_q}} m_{K^*} . \eqno(4)
$$
This obviously yields usual GMO formula [10]
$3m_{{\omega}_8} + m_{\rho} = 4 m_{K^*}$
(known to require manifest introducing of singlet mixing)
at $q=1$ (in this case $[2]_q/[3]_q=2/3$), but also produces
the formula
$$ m_{{\omega}_8} + m_{\rho} = 2 m_{K^*}      \hskip 14mm
    {\rm at} \qquad \quad [2]_q=[3]_q                \eqno(5)
$$
(the latter holds if $q=q_3 = e^{i\pi /5}$). With
$m_{{\omega}_8}\equiv m_{\phi}$, Eq.(5) coincides with
nonet mass formula of Okubo [12] which shows
\underline {ideal agreement} with data [13] (up to errors of
experiment and of averaging over isoplets).
What are higher analogues of Okubo's sum rule?
From (3), with natural choice
$[n]_q=[n-1]_q$, $n=4,5,6,$ we get them:
\vskip -1mm
$$ m_{{\omega}_{15}} + (5 - 8/[2]_{q_4}) m_{\rho} =
 2\ m_{D^*} + (4 - 8/[2]_{q_4}) m_{K^*}   \hskip 38mm    \eqno(6)
$$
$$ m_{{\omega}_{24}} + (9 - 16/[2]_{q_5}) m_{\rho} =
2\ m_{{D_b}^*} + (4 - 8/[2]_{q_5})(m_{D^*}
                             + m_{K^*})   \hskip 18mm   \eqno(7)
$$
$$
 m_{{\omega}_{35}} + (13 - 24/[2]_{q_6}) m_{\rho} =
2\ m_{{D_t}^*}
+ (4 - 8/[2]_{q_6})(m_{{D_b}^*} + m_{D^*} + m_{K^*}).  \eqno(8)
$$
\vskip 1mm
\ni Here $q_n$ denote the values that solve
eqns. $\ [n]_q-[n-1]_q=0$, namely,
$$
q_n = e^{i\pi /(2n-1)}.                             \eqno(9)
$$
Like in the case with $m_{{\omega}_8}\equiv m_{\phi},$ it is meant
in (6) that $J/\psi $ is put in place of ${\omega }_{15},$ etc.

The roots (9) and their 'master' polynomials (in $q$) have a
topological implication.

\bigskip
\ni{\bf 3. Knot structures associated with vector quarkonia}
\bigskip

    The quantities $[n]_q-[n - 1]_q,$ by their roots,
reduce the $q$-analogues (3) to realistic mass sum rules (5),
(6)-(8). At the same time, being such polynomials
$P_n(q)$ that satisfy (see [14])
$$
(i)\ \ P_n(q)=P_n(q^{-1}),              \hskip 26mm
           (ii)\ \ P_n(1)=1,
$$
they do coincide with topological invariants -
Alexander polynomials $\Delta (q)\{(2n-1)_1\}$ of torus $(2n-1)_1$-knots.
For instance,
$$[3]_q-[2]_q=q^2+q^{-2}-q-q^{-1}+1\equiv \Delta (q)\{ 5_1\} ,
                                       \hskip 20mm      \eqno(10)
$$
$$
[4]_q-[3]_q=
q^3+q^{-3}-q^2-q^{-2}+q+q^{-1}-1\equiv \Delta (q)\{ 7_1\}, \eqno(11)
$$
correspond to the $5_1$- and $7_1$-
knots, see fig. 1.
The 'extra' $q$-deuce in (3) may be related with
the trefoil (or $3_1$-) knot, since
$[2]_q -1=q+q^{-1}-1\equiv\Delta (q)\{3_1\}$. Hence,
{\it all the} $q$-{\it dependence} in masses of $\omega_{n^2-1}$
and in coefficients of Eq.(3) is expressible
through various Alexander polynomials:
$$
{[3]_q\over [2]_q }=1+{\Delta \{5_1\}\over [2]_q}=
1+{\Delta \{5_1\}\over \Delta \{3_1\} +1},  \hskip 60mm \eqno (12)
$$
$$
{[n]_q\over [n-1]_q }=
1+{\Delta \{ (2n-1)_1\}\over [n-1]_q}=1+{\Delta \{ (2n-1)_1\}\over
1+ \sum_{r=2}^{n-1} \Delta \{(2r-1)_1\} },  \qquad       n=4,5,6.
                                                       \eqno (13)
$$
   Thus, the values (9) are roots of respective Alexander
polynomials. For every fixed $n$, it is the 'senior' polynomial
in Eqs.(12),(13) (the one in numerator)  which is
distinguished: it serves to extract, through its root,
the corresponding MSR from $q$-deformed analogue. At $n=3$
the $q$-GMO formula (4) generates simple and successful Okubo's
relation (5); at $n=4,5,6, $ general formula (3) reduces
to the higher analogues (6)-(8) of Okubo's relation.

Let us emphasize that, for particular $n,$  the corresponding
value $q_n$ is fixed in a rigid way as a root of
$\Delta \{ (2n-1)_1\},$ contrary to a choice of $q$ by fitting in
other phenomenological applications [3]. For instance,
\underline {rigid fixation} $q={\rm e}^{i\pi /5}$ turns
$\Delta \{5_1\}$ into zero as well as the ratio $[3]_q/[2]_q$
into unity, thus extracting (5) from the $q$-analogue (4).
This extends to higher $n=4,5,6.$

\vspace{-5mm}
\special{em:linewidth 0.4pt}
\unitlength 1.00mm
\linethickness{0.4pt}
\begin{picture}(136.33,81.67)
\emline{11.67}{30.33}{1}{11.28}{27.94}{2}
\emline{11.28}{27.94}{3}{11.02}{25.75}{4}
\emline{11.02}{25.75}{5}{10.89}{23.76}{6}
\emline{10.89}{23.76}{7}{10.90}{21.96}{8}
\emline{10.90}{21.96}{9}{11.05}{20.36}{10}
\emline{11.05}{20.36}{11}{11.33}{18.96}{12}
\emline{11.33}{18.96}{13}{11.75}{17.75}{14}
\emline{11.75}{17.75}{15}{12.30}{16.75}{16}
\emline{12.30}{16.75}{17}{12.99}{15.94}{18}
\emline{12.99}{15.94}{19}{13.81}{15.32}{20}
\emline{13.81}{15.32}{21}{14.77}{14.91}{22}
\emline{14.77}{14.91}{23}{15.86}{14.69}{24}
\emline{15.86}{14.69}{25}{17.08}{14.66}{26}
\emline{17.08}{14.66}{27}{18.45}{14.84}{28}
\emline{18.45}{14.84}{29}{19.94}{15.21}{30}
\emline{19.94}{15.21}{31}{21.58}{15.78}{32}
\emline{21.58}{15.78}{33}{23.34}{16.55}{34}
\emline{23.34}{16.55}{35}{25.25}{17.51}{36}
\emline{25.25}{17.51}{37}{27.28}{18.68}{38}
\emline{27.28}{18.68}{39}{29.46}{20.03}{40}
\emline{29.46}{20.03}{41}{31.76}{21.59}{42}
\emline{31.76}{21.59}{43}{34.21}{23.34}{44}
\emline{34.21}{23.34}{45}{36.78}{25.29}{46}
\emline{36.78}{25.29}{47}{39.50}{27.44}{48}
\emline{39.50}{27.44}{49}{42.35}{29.79}{50}
\emline{42.35}{29.79}{51}{45.33}{32.33}{52}
\emline{31.00}{18.33}{53}{33.06}{17.13}{54}
\emline{33.06}{17.13}{55}{35.00}{16.13}{56}
\emline{35.00}{16.13}{57}{36.80}{15.33}{58}
\emline{36.80}{15.33}{59}{38.48}{14.73}{60}
\emline{38.48}{14.73}{61}{40.02}{14.32}{62}
\emline{40.02}{14.32}{63}{41.44}{14.11}{64}
\emline{41.44}{14.11}{65}{42.73}{14.10}{66}
\emline{42.73}{14.10}{67}{43.88}{14.29}{68}
\emline{43.88}{14.29}{69}{44.91}{14.67}{70}
\emline{44.91}{14.67}{71}{45.81}{15.26}{72}
\emline{45.81}{15.26}{73}{46.58}{16.04}{74}
\emline{46.58}{16.04}{75}{47.22}{17.02}{76}
\emline{47.22}{17.02}{77}{47.74}{18.19}{78}
\emline{47.74}{18.19}{79}{48.12}{19.57}{80}
\emline{48.12}{19.57}{81}{48.37}{21.14}{82}
\emline{48.37}{21.14}{83}{48.50}{22.91}{84}
\emline{48.50}{22.91}{85}{48.49}{24.88}{86}
\emline{48.49}{24.88}{87}{48.36}{27.05}{88}
\emline{48.36}{27.05}{89}{48.09}{29.41}{90}
\emline{48.09}{29.41}{91}{47.70}{31.98}{92}
\emline{47.70}{31.98}{93}{47.18}{34.74}{94}
\emline{47.18}{34.74}{95}{46.52}{37.70}{96}
\emline{46.52}{37.70}{97}{45.74}{40.85}{98}
\emline{45.74}{40.85}{99}{44.83}{44.21}{100}
\emline{44.83}{44.21}{101}{43.00}{50.33}{102}
\emline{22.67}{53.33}{103}{26.27}{53.49}{104}
\emline{26.27}{53.49}{105}{29.65}{53.56}{106}
\emline{29.65}{53.56}{107}{32.81}{53.56}{108}
\emline{32.81}{53.56}{109}{35.75}{53.47}{110}
\emline{35.75}{53.47}{111}{38.48}{53.31}{112}
\emline{38.48}{53.31}{113}{40.99}{53.06}{114}
\emline{40.99}{53.06}{115}{43.28}{52.74}{116}
\emline{43.28}{52.74}{117}{45.35}{52.33}{118}
\emline{45.35}{52.33}{119}{47.21}{51.85}{120}
\emline{47.21}{51.85}{121}{48.85}{51.28}{122}
\emline{48.85}{51.28}{123}{50.27}{50.63}{124}
\emline{50.27}{50.63}{125}{51.47}{49.90}{126}
\emline{51.47}{49.90}{127}{52.46}{49.10}{128}
\emline{52.46}{49.10}{129}{53.23}{48.21}{130}
\emline{53.23}{48.21}{131}{53.78}{47.24}{132}
\emline{53.78}{47.24}{133}{54.12}{46.19}{134}
\emline{54.12}{46.19}{135}{54.23}{45.06}{136}
\emline{54.23}{45.06}{137}{54.13}{43.85}{138}
\emline{54.13}{43.85}{139}{53.81}{42.56}{140}
\emline{53.81}{42.56}{141}{53.28}{41.19}{142}
\emline{53.28}{41.19}{143}{52.52}{39.74}{144}
\emline{52.52}{39.74}{145}{51.55}{38.20}{146}
\emline{51.55}{38.20}{147}{49.33}{35.33}{148}
\emline{26.67}{20.33}{149}{23.65}{22.68}{150}
\emline{23.65}{22.68}{151}{20.86}{24.94}{152}
\emline{20.86}{24.94}{153}{18.28}{27.11}{154}
\emline{18.28}{27.11}{155}{15.92}{29.20}{156}
\emline{15.92}{29.20}{157}{13.78}{31.20}{158}
\emline{13.78}{31.20}{159}{11.86}{33.11}{160}
\emline{11.86}{33.11}{161}{10.16}{34.94}{162}
\emline{10.16}{34.94}{163}{8.68}{36.68}{164}
\emline{8.68}{36.68}{165}{7.42}{38.33}{166}
\emline{7.42}{38.33}{167}{6.38}{39.89}{168}
\emline{6.38}{39.89}{169}{5.56}{41.37}{170}
\emline{5.56}{41.37}{171}{4.95}{42.76}{172}
\emline{4.95}{42.76}{173}{4.57}{44.06}{174}
\emline{4.57}{44.06}{175}{4.41}{45.28}{176}
\emline{4.41}{45.28}{177}{4.46}{46.41}{178}
\emline{4.46}{46.41}{179}{4.74}{47.45}{180}
\emline{4.74}{47.45}{181}{5.23}{48.41}{182}
\emline{5.23}{48.41}{183}{5.95}{49.28}{184}
\emline{5.95}{49.28}{185}{6.88}{50.06}{186}
\emline{6.88}{50.06}{187}{8.04}{50.75}{188}
\emline{8.04}{50.75}{189}{9.41}{51.36}{190}
\emline{9.41}{51.36}{191}{11.00}{51.88}{192}
\emline{11.00}{51.88}{193}{12.81}{52.32}{194}
\emline{12.81}{52.32}{195}{14.84}{52.66}{196}
\emline{14.84}{52.66}{197}{18.00}{53.00}{198}
\emline{13.00}{35.00}{199}{14.15}{38.63}{200}
\emline{14.15}{38.63}{201}{15.29}{42.02}{202}
\emline{15.29}{42.02}{203}{16.43}{45.18}{204}
\emline{16.43}{45.18}{205}{17.57}{48.10}{206}
\emline{17.57}{48.10}{207}{18.70}{50.79}{208}
\emline{18.70}{50.79}{209}{19.83}{53.23}{210}
\emline{19.83}{53.23}{211}{20.95}{55.44}{212}
\emline{20.95}{55.44}{213}{22.07}{57.42}{214}
\emline{22.07}{57.42}{215}{23.19}{59.16}{216}
\emline{23.19}{59.16}{217}{24.30}{60.66}{218}
\emline{24.30}{60.66}{219}{25.40}{61.92}{220}
\emline{25.40}{61.92}{221}{26.50}{62.95}{222}
\emline{26.50}{62.95}{223}{27.60}{63.74}{224}
\emline{27.60}{63.74}{225}{28.69}{64.30}{226}
\emline{28.69}{64.30}{227}{29.78}{64.61}{228}
\emline{29.78}{64.61}{229}{30.87}{64.70}{230}
\emline{30.87}{64.70}{231}{31.95}{64.54}{232}
\emline{31.95}{64.54}{233}{33.02}{64.15}{234}
\emline{33.02}{64.15}{235}{34.09}{63.52}{236}
\emline{34.09}{63.52}{237}{35.16}{62.66}{238}
\emline{35.16}{62.66}{239}{36.22}{61.56}{240}
\emline{36.22}{61.56}{241}{37.28}{60.22}{242}
\emline{37.28}{60.22}{243}{38.34}{58.64}{244}
\emline{38.34}{58.64}{245}{39.39}{56.83}{246}
\emline{39.39}{56.83}{247}{40.33}{55.00}{248}
\emline{122.00}{13.33}{249}{119.07}{11.89}{250}
\emline{119.07}{11.89}{251}{116.74}{10.44}{252}
\emline{116.74}{10.44}{253}{115.00}{9.00}{254}
\emline{115.00}{9.00}{255}{113.85}{7.56}{256}
\emline{113.85}{7.56}{257}{113.29}{6.11}{258}
\emline{113.29}{6.11}{259}{113.33}{4.67}{260}
\emline{116.67}{68.67}{261}{114.44}{69.82}{262}
\emline{114.44}{69.82}{263}{112.75}{71.25}{264}
\emline{112.75}{71.25}{265}{111.61}{72.98}{266}
\emline{111.61}{72.98}{267}{111.00}{75.00}{268}
\emline{119.67}{38.67}{269}{117.43}{37.61}{270}
\emline{117.43}{37.61}{271}{115.60}{36.56}{272}
\emline{115.60}{36.56}{273}{114.16}{35.53}{274}
\emline{114.16}{35.53}{275}{113.14}{34.51}{276}
\emline{113.14}{34.51}{277}{112.51}{33.51}{278}
\emline{112.51}{33.51}{279}{112.30}{32.53}{280}
\emline{112.30}{32.53}{281}{112.48}{31.56}{282}
\emline{112.48}{31.56}{283}{113.07}{30.61}{284}
\emline{113.07}{30.61}{285}{114.06}{29.67}{286}
\emline{114.06}{29.67}{287}{115.46}{28.75}{288}
\emline{115.46}{28.75}{289}{117.67}{27.67}{290}
\emline{120.67}{26.00}{291}{118.34}{24.97}{292}
\emline{118.34}{24.97}{293}{116.40}{23.93}{294}
\emline{116.40}{23.93}{295}{114.87}{22.90}{296}
\emline{114.87}{22.90}{297}{113.72}{21.86}{298}
\emline{113.72}{21.86}{299}{112.98}{20.83}{300}
\emline{112.98}{20.83}{301}{112.63}{19.79}{302}
\emline{112.63}{19.79}{303}{112.68}{18.76}{304}
\emline{112.68}{18.76}{305}{113.12}{17.72}{306}
\emline{113.12}{17.72}{307}{113.96}{16.69}{308}
\emline{113.96}{16.69}{309}{115.20}{15.66}{310}
\emline{115.20}{15.66}{311}{118.00}{14.00}{312}
\emline{124.33}{38.33}{313}{126.09}{37.26}{314}
\emline{126.09}{37.26}{315}{127.45}{36.18}{316}
\emline{127.45}{36.18}{317}{128.41}{35.09}{318}
\emline{128.41}{35.09}{319}{128.96}{34.00}{320}
\emline{128.96}{34.00}{321}{129.10}{32.91}{322}
\emline{129.10}{32.91}{323}{128.85}{31.81}{324}
\emline{128.85}{31.81}{325}{128.19}{30.70}{326}
\emline{128.19}{30.70}{327}{127.12}{29.59}{328}
\emline{127.12}{29.59}{329}{125.65}{28.47}{330}
\emline{125.65}{28.47}{331}{123.78}{27.35}{332}
\emline{123.78}{27.35}{333}{121.00}{26.00}{334}
\emline{122.67}{13.33}{335}{124.87}{14.51}{336}
\emline{124.87}{14.51}{337}{126.62}{15.69}{338}
\emline{126.62}{15.69}{339}{127.91}{16.87}{340}
\emline{127.91}{16.87}{341}{128.74}{18.06}{342}
\emline{128.74}{18.06}{343}{129.12}{19.24}{344}
\emline{129.12}{19.24}{345}{129.05}{20.42}{346}
\emline{129.05}{20.42}{347}{128.51}{21.60}{348}
\emline{128.51}{21.60}{349}{127.53}{22.78}{350}
\emline{127.53}{22.78}{351}{125.00}{24.67}{352}
\emline{120.00}{39.00}{353}{122.37}{40.04}{354}
\emline{122.37}{40.04}{355}{124.40}{41.08}{356}
\emline{124.40}{41.08}{357}{126.06}{42.13}{358}
\emline{126.06}{42.13}{359}{127.38}{43.17}{360}
\emline{127.38}{43.17}{361}{128.33}{44.21}{362}
\emline{128.33}{44.21}{363}{128.94}{45.25}{364}
\emline{128.94}{45.25}{365}{129.19}{46.29}{366}
\emline{129.19}{46.29}{367}{129.09}{47.33}{368}
\emline{129.09}{47.33}{369}{128.63}{48.37}{370}
\emline{128.63}{48.37}{371}{127.82}{49.42}{372}
\emline{127.82}{49.42}{373}{126.66}{50.46}{374}
\emline{126.66}{50.46}{375}{123.67}{52.33}{376}
\emline{125.33}{11.33}{377}{127.54}{9.66}{378}
\emline{127.54}{9.66}{379}{128.98}{8.00}{380}
\emline{128.98}{8.00}{381}{129.67}{5.33}{382}
\emline{117.67}{40.33}{383}{115.74}{41.39}{384}
\emline{115.74}{41.39}{385}{114.19}{42.44}{386}
\emline{114.19}{42.44}{387}{113.02}{43.48}{388}
\emline{113.02}{43.48}{389}{112.22}{44.52}{390}
\emline{112.22}{44.52}{391}{111.80}{45.55}{392}
\emline{111.80}{45.55}{393}{111.75}{46.58}{394}
\emline{111.75}{46.58}{395}{112.08}{47.61}{396}
\emline{112.08}{47.61}{397}{112.78}{48.63}{398}
\emline{112.78}{48.63}{399}{113.85}{49.65}{400}
\emline{113.85}{49.65}{401}{115.31}{50.66}{402}
\emline{115.31}{50.66}{403}{117.13}{51.66}{404}
\emline{117.13}{51.66}{405}{119.33}{52.67}{406}
\emline{120.33}{53.00}{407}{122.68}{54.17}{408}
\emline{122.68}{54.17}{409}{124.66}{55.33}{410}
\emline{124.66}{55.33}{411}{126.28}{56.47}{412}
\emline{126.28}{56.47}{413}{127.53}{57.59}{414}
\emline{127.53}{57.59}{415}{128.42}{58.70}{416}
\emline{128.42}{58.70}{417}{128.94}{59.78}{418}
\emline{128.94}{59.78}{419}{129.10}{60.85}{420}
\emline{129.10}{60.85}{421}{128.89}{61.91}{422}
\emline{128.89}{61.91}{423}{128.33}{62.94}{424}
\emline{128.33}{62.94}{425}{127.39}{63.96}{426}
\emline{127.39}{63.96}{427}{126.09}{64.96}{428}
\emline{126.09}{64.96}{429}{123.67}{66.33}{430}
\emline{118.33}{54.33}{431}{116.33}{55.32}{432}
\emline{116.33}{55.32}{433}{114.68}{56.30}{434}
\emline{114.68}{56.30}{435}{113.40}{57.28}{436}
\emline{113.40}{57.28}{437}{112.48}{58.26}{438}
\emline{112.48}{58.26}{439}{111.92}{59.23}{440}
\emline{111.92}{59.23}{441}{111.73}{60.20}{442}
\emline{111.73}{60.20}{443}{111.89}{61.16}{444}
\emline{111.89}{61.16}{445}{112.42}{62.12}{446}
\emline{112.42}{62.12}{447}{113.31}{63.07}{448}
\emline{113.31}{63.07}{449}{114.57}{64.03}{450}
\emline{114.57}{64.03}{451}{116.18}{64.97}{452}
\emline{116.18}{64.97}{453}{120.00}{66.67}{454}
\emline{120.33}{67.00}{455}{123.34}{68.61}{456}
\emline{123.34}{68.61}{457}{125.68}{70.12}{458}
\emline{125.68}{70.12}{459}{127.37}{71.52}{460}
\emline{127.37}{71.52}{461}{128.39}{72.82}{462}
\emline{128.39}{72.82}{463}{128.67}{74.67}{464}
\end{picture}

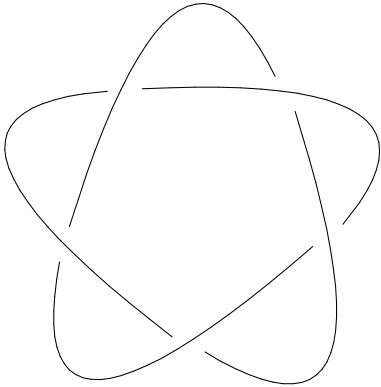
\small {Fig. 1.\   Torus $5_1$-knot
      corresponding to $[3]_q-[2]_q$    \hskip 12mm
        Fig. 2.\    Braid with 5 overcrossings }

\normalsize
\bigskip
\medskip
Using flavor $q$-algebras along with 'dynamical' $q$-algebras
(through embedding $U_q(u_{n})\subset U_q(u_{n+1})$
we get as result that the collection of torus knots
$5_1,\ 7_1,\ 9_1,\ 11_1$ \ is put into
correspondence [7] with vector quarkonia
$s\bar s$, $c\bar c$, $b\bar b$, and $t\bar t$ respectively.
Thus, application of the $q$-algebras suggests a possibility of
{\it topological characterization of flavors}: fixed number $n$ just
corresponds to  $2n\!-\!1$  overcrossings of 2-strand braids
(see fig. 2) whose
closures give these $(2n-1)_1$-torus knots. Or, with the form
$(2n-1,2)$ of these same knots, we have the correspondence
$n \leftarrow\!\rightarrow w\equiv 2n-1,\ $
where $w$  means the winding number around the body (tube)
of torus (winding number around the hole of torus being equal to 2
for all $n\ge 3$).

In other words, to compare with empirical data, one has to
fix appropriately the parameter $q$: it appears that to
each number $n,\ n\ge 3,$ there corresponds a prime
root of unity $q=q(n)=e^{{i\pi}/{(2n-1)}}$.
The latter turns into zero the polynomial
$P_{n}(q)\equiv [n]_q-[n-1]_q$ that coincides with respective
Alexander polynomial of the torus $(2n-1)_1$-knot [4,7].
In a sense, the polynomial $P_{n}(q)$ through
its root $q(n)$ determines the strength of deformation at every
fixed $n$ and, due to this, serves as
{\it defining polynomial} for the corresponding mass sum rule
(and quarkonium).

Let us remark that a (purely heuristic) picture which assigns
knot-like structures to some of fundamental particles was
proposed in [15].

\newpage

\ni {\bf 4. Octet baryon mass formulas: $q$-deformation
               lifts 'degeneracy'}
\bigskip

The approach similar to that in [4] was applied to baryons
${1\over 2}^+$,
including charmed ones, by adopting again $U_q(u_4)$
for the $4$-flavor symmetry.
However, unlike the case of vector mesons treated with
'compact' $q$-algebras $U_q(u_{n+1})$ of dynamical symmetry,
here the irreps of 'noncompact' dynamical symmetry,
realized by the $q$-algebra ${\cal A}\equiv U_q(u_{4,1}),$
were first exploited [5].

   For the standard GMO sum rule [10] for baryon octet masses
$$m_N+m_{\Xi}={3\over 2}m_{\Lambda}+{1\over 2}m_{\Sigma }  \eqno (14)
$$
known to hold with $0.58 \%$ accuracy,
a $q$-deformed analogue was derived [5,7] which contains
(14) as its classical $q=1$ limit. This was first performed
within the concrete irrep\footnote{  Necessary details concerning
irreps of the algebra $U_q(u_{4,1})$ are given in [5,7] too.}
${\cal D}_0\equiv D^+_{12}(p-1,p-\!3,p-4;p,p-2)$ ($p$ is some fixed
integer not entering final results). Note that
for state vectors of octet baryons, embedded together with entire
$20$-plet of $U_q(su_4)$ into this dynamical irrep of $\cal{A},$
the Gel'fand-Zetlin basis vectors are applied;
mass operator in its $U_q(su_3)$-  and $U_q(su_4)$-breaking terms
involves bilinearly those representation operators of ${\cal D}_0$
which are extra as regards the 'compact' subalgebra $U_q(su_4)$.

Evaluating octet baryon masses
within this dynamical irrep ${\cal D}_0$
results in the $q$-analogue
$$
[2]\, m_N+{[2]\over [2]-1}\, m_{\Xi }={[3]}\, m_{\Lambda }
+\Bigl ({[2]^2\over [2]-1}-{[3]}\Bigr )\, m_{\Sigma }
  +{A_q\over B_q}\ {\cal C}_{\rm mass}                   \eqno(15)
$$
where
${\cal C}_{\rm mass}
      = m_{\Xi } - m_{\Lambda } - [2] ( m_{\Sigma } - m_N ),$
$$
 A_q = ([2]-2)[2]^4([2]^2-3),            \hspace{14 mm}
   B_q =\bigl ([2]^3-[2]^2[4]+3[5]-
                    [3]\bigr )([2]-1).                  \eqno(16)
$$
In this $q$-deformed mass relation, most significant
is the remarkable rightmost term with $A_q/B_q$ as prefactor.
Due to $A_q ,$  this term vanishes for some values of $q$
including the 'classical' one $q=1.$
The polynomial $A_q$, by its zeros, {\it determines} both the GMO
(at $q=1$, i.e. $[2]_q=2$) and, at $q\neq 1 ,$ some other mass sum rules.
That is, $A_q$  plays the key role: {\it it puts on equal footing} the
GMO sum rule and some others. Namely,

at $q=1,\ $ Eq.(15) reduces to the standard GMO sum rule (14);

at $q= e^{i\pi /2}$ (now $[2]_q=0$) we get the equality
$m_{\Sigma }=m_{\Lambda }$
(rough one, empirically);

at $q= e^{i\pi / 6}$  (then $[2]_{q}=\pm \sqrt{3}$ )
the relation (15) yields new mass sum rule
$$
m_N + {1\!+\!\sqrt 3\over 2}\ m_{\Xi} = {2\over{\sqrt 3}}\ m_{\Lambda }
         + {9\!-\!\sqrt 3\over 6}\ m_{\Sigma } .         \eqno{(17)}
$$
Comparison with data for baryon masses [13] shows the precision
of $0.22 \%$ ($2739.5 $ MeV versus $ 2733.4\ $ MeV) - this is
significantly better than in the case of GMO.

\bigskip

Such possibility to obtain new mass sum rules which are
theoretically on equal footing with the GMO one, inspires to
search for dynamical representations, either of the
'noncompact' $U_q(u_{4,1})$ or the compact $U_q(u_{5})$
version of ($q$-deformed) dynamical symmetry,
capable to yield relations like (15) but with differing sets
of zeroes for relevant $A_q.$
And, what is really surprising, the new sum rule (17) obtained
within the specific dynamical irrep ${\cal D}_0$ of $U_q(u_{4,1})$
is still not the best one.
It can be proved [16], using
the compact dynamical $U_q(u_5),$ that among the admissible
dynamical irreps there exist an entire series
(labelled by an integer $k,\ $ $0\le k<\infty $) of irreps capable
to produce even more accurate MSR.

\medskip
{\sl Proposition.} An infinite series of $q$-deformed relations
of the form (15), labelled by an integer $n,$ can be obtained which
differ in their defining polynomials: in $n$-th relation, the
quantity $A_q=A_q(n)$ has, besides the two obligue roots $q=1$
and $q=e^{i\pi /2}$ (i.e., $[2]=0$), the additional root $q=e^{i\pi /n}$
exhibited by $[n]=0$. The set of roots of the corresponding
$B_q=B_q(n)$ has zero intersection with that of $A_q(n)$.

This can be shown by pointing out those concrete dynamical
irreps of $U_q(u_5)$ which, after computations,
 yield just the relations mentioned
in the proposition.

To find most optimal choise it is now sufficient to analyse agreement
of (15) (at vanishing term ${A_q\over B_q}\ {\cal C}_{\rm mass}$)
with data, using a kind of {\it 'discrete' fitting procedure}.
The results are shown in the table.

\begin{center}
\begin{tabular}{|c|c|c|c|c|}
\hline
$\t=\frac\pi n$ & LHS,\ \ $MeV$ & RHS,\ \ $MeV$
    & (RHS $-$LHS),\ \ $MeV$
   & ${\left | {\rm RHS}-{\rm LHS}\right |\over {\rm RHS} }, \ \  \%$  \\
\hline
$\pi /\infty$ & 4514.0 & 4540.2 & 26.2 & 0.58  \\
$\pi /30$ & 4518.31 & 4543.73 & 25.42 & 0.56  \\
$\pi /12$ & 4546.41 & 4566.61 & 20.2 & 0.44  \\
$\pi /9$ & 4581.54 & 4595.9  & 14.36 & 0.31  \\
$\pi /8$ & 4607.77 & 4618.16 & 10.39 & 0.23  \\
$\pi /7$ & 4653.58 & 4656.85 & 3.26 & 0.07  \\
$\pi /6$ & 4744.88 & 4734.41 & -10.47 & 0.22  \\
$\pi /5$ & 4970.0 & 4928.82 & -41.18 & 0.84  \\
\hline
\end{tabular}
\end{center}

\medskip
\ni This table clearly reflects the fact of existence of infinite discrete
set of mass formulas labelled by an integer $n\ (6\le n < \infty),$
namely
$$
m_N+{1\over [2]_{q_n}-1}m_{\Xi }=
{[3]_{q_n}\over [2]_{q_n}}m_{\Lambda }
+\Bigl ({[2]_{q_n}\over [2]_{q_n}-1}-
{[3]_{q_n}\over [2]_{q_n}}\Bigr )m_{\Sigma },
                                                      \eqno(18)
$$
{\em each of which shows better agreement with data} than the
classical GMO one (here $q_n=e^{{\rm i}\pi/n}$).

Now it is easy to see: \underline {the best choice} corresponds to
$q=q_7=e^{i\t_{\bf 8} },\ \t_{\bf 8}\equiv\pi /7.$
\underline {In this case}
$$
[7]_{q_7}=0,  \hspace{30mm}
[3]_{q_7} = {[2]_{q_7}\over [2]_{q_7}-1},
$$
and we get the new MSR in the form
$$
m_N + {m_{\Xi}\over [2]_{q_7}-1 } =
{m_{\Lambda}\over [2]_{q_7}-1 } +  m_{\Sigma }       \eqno{(19)}
$$
which shows ($2582.6 $ MeV versus $2584.4 $ MeV)
remarkable $0.07 \%$ accuracy!
Moreover, such excellent precision is combined with an
apparent simplicity: equality of the coefficients
at $m_N$ and $m_{\Sigma},$ as well as those
at $m_{\Xi}$ and $m_{\Lambda}$. Due to that, Eq.(19)
takes equivalent form
(recall that $[2]_{q_7}=q_7+(q_7)^{-1}=2 \cos \frac\pi 7$)
$$
 m_{\Xi}-m_{N}+m_{\Sigma}-m_{\Lambda} =
(2\cos \frac\pi 7)(m_{\Sigma}-m_{N})                  \eqno(20)
$$
which is of interest being very similar to the
decuplet mass formula, see Eq.(21) below.

\bigskip
\ni {\bf 5. Decuplet baryon masses: essentially nonlinear
$SU(3)$-breaking effects}
\bigskip

In the case of baryons ${3\over 2}^+$ from the $SU(3)$ decuplet
it is known that convensional (first order) symmetry breaking
yields equal spacing rule (ESR) for masses of isoplet members in
${\bf 10}$-plet [10]. Empirical data show that actually there is
noticeable deviation from ESR:
$$
   m_{\Sigma^*}-m_{\Delta}\hskip 26mm
   m_{\Xi^*}-m_{\Sigma^*}\hskip 26mm
   m_{\Omega}-m_{\Xi^*}
$$
$$
  152.6\ MeV  \hskip 26mm   148.8\ MeV \hskip 26mm  139.0\ MeV
$$
It was shown in [6] (in analogy to octet case) that use of the
$q$-algebras $U_q(su_n),$ instead of $SU(n),$ provides natural
and simple improvement of situation. From evaluations
of decuplet masses
in two particular irreps of the dynamical
$U_q(u_{4,1})$, the $q$-deformed mass relation
$$
(1/[2]_q)(m_{\Sigma^*}-m_{\Delta}+m_{\Omega}-m_{\Xi^*})
= m_{\Xi^*}-m_{\Sigma^*}, \hskip 14mm [2]_q\equiv q+q^{-1}, \eqno(21)
$$
was obtained. As proven in [6], this mass relation
is a {\it universal one} - it follows within any admissible
irrep (such that contains $SU_q(3)$-decuplet embedded
in ${\sl 20}$-plet of $SU_q(4)$)
of the dynamical $U_q(u_{4,1})$.
Taking empirical masses  [13] $m_{\Delta}=1232$ MeV,
\ $m_{\Sigma^*}=1384.6$ MeV, \ $m_{\Xi^*}=1533.4$ MeV ,  and
$m_{\Omega}=1672.4$ MeV  we see that the formula (21), to be
successfull, requires $[2]_q\simeq 1.96$ . But this never holds
for real $q$. On the contrary, \underline {pure phase} $q=e^{i\t}$
(in this case $[2]_q=2\cos\t$) with fixed
$\t=\t_{\bf 10}\simeq\frac{\pi }{14}$
provides remarkable agreement with data.

\smallskip
Observe the apparent similarity of eq.(21) with mass relation
$$
(1/2)(m_{\Sigma^{*}}-m_{\Delta}+m_{\Omega}-m_{\Xi^{*}})
= m_{\Xi^{*}}-m_{\Sigma^{*}}                           \eqno(22)
$$
obtained earlier in different contexts: within tensor method
of eightfold way [17]; in additive quark model with most general
quark--quark pair interaction [18]; within the diquark-quark model
of [19]; in the framework of modern chiral perturbation theory [20].
Such almost model-independence of (22) is based on the fact that
each of these approaches takes into account in its specific way
not only first, but also the 2nd order of $SU(3)$ symmetry breaking.

\smallskip
Universality of the $q$-deformed MSR (21) holds even in a
wider sense: it extends to all admissible irreps
(also containing ${\sl 20}$-plet of $SU_q(4)$) of the
'compact' dynamical $SU_q(5)$ .
Let us exemplify the decuplet masses by
taking simplest although typical instance of irrep.
Say, within most symmetric dynamical irrep
$\{ 4 0 0 0 \}$ of $SU_q(5)$ calculation yields
$m_{\D}=m_{\bf 10} +  \b ,\ $
$m_{\S^*}=m_{\bf 10} + [2] \b +  \a ,\ $
$m_{\Xi^*}=m_{\bf 10} + [3] \b + [2] \a ,\ $
$m_{\O}=m_{\bf 10} + [4] \b + [3] \a ,\ $
and these obviously satisfy (21).

All these masses can be comprised by the formula with
explicit dependence on hypercharge:
$$
m_B = m\bigl( Y(B) \bigr ) =
m_{\bf 10} + [1\!-\!Y]\ {\a} + [2\!-\!Y]\ {\b}.             \eqno(23)
$$
In the $q=1$ limit it reduces to familiar formula
$$
m_B = \tilde{m}_{\bf 10} + a\ Y                            \eqno(24)
$$
with {\it linear dependence on hypercharge} $Y$
(or strangeness); here $ a = - \a - \b ,\ $
$\tilde{m}_{\bf 10} = m_{\bf 10} + \a + 2 \b.$
On the contrary, one can see that formula (23)
{\it involves highly nonlinear dependence} of mass on hypercharge
(it is $Y$ alone which causes $SU(3)$ breaking in the decuplet case).
Indeed, for $q$-number $[N]$ we have
$[N]=q^{N-1}+q^{N-3}+\ldots +q^{-N+3}+q^{-N+1},\ $
($N$ terms in total) that visualizes essential $Y$-nonlinearity of
(23). Here is the principal difference between (21) and (22):
as already noted, the latter accounts only terms which are
linear and quadratic in $Y.$

\bigskip
\ni{\bf 6. Baryon masses and the Cabibbo angle}
\bigskip

As it was found in the case of mesons, the parameter $q,$ being pure
phase at any $n,$ is closely related with the issue of (singlet)
mixing. Below we argue that $q$ is connectible with the
fundamental mixing angle encountered in particle theory -
the Cabibbo angle $\t_{\rm C}$.

It is known for a long time that mass relations may involve
the Cabibbo angle.  At the constituent (quark) level this is
exhibited by the relation [11]
$$
\tan^2\theta_{\rm C} = {m_d}/{m_s},                           \eqno(25)
$$
while at the composite level of pseudoscalar mesons
this is seen, e.g., from the formula [21]
$$
\vspace{-1mm}
\tan^2\theta_{\rm C} = \frac{ m^2_{\pi} }
        { m^2_{K}\frac{F_K}{F_\pi} -m^2_{\pi} } ,
\vspace{-2mm}
$$
or from Weinberg's formula [22]
$$
\vspace{-2mm}
\frac{m_d}{m_s}=
\frac{m^2_{K^0}+m^2_{\pi^+}-m^2_{K^+}}
     {m^2_{K^0}-m^2_{\pi^+}+m^2_{K^+}}
$$
combined with (25). There exists even wider variety of formulas
involving, besides meson masses from the octet $0^-$,
some {\it additional dimensionless parameter(s)}
such as $F_K/F_{\pi}$, together with (or instead of) $\t_{\rm C}$.
Among these,  most relevant for us is the relation [23]
$$
m_\pi^2 + 3 \frac{F_\eta^2 m_\eta^2}{F_\pi^2}
= 4 \frac{F_K^2 m_K^2}{F_\pi^2}                      \eqno(26)
$$
which generalizes the GMO mass formula.
This is to be compared with pseudoscalar version
(involving masses squared) of our $q$-deformed
meson mass relation (4):
$$
m_\pi^2 + \frac{[3]}{2 [2] - [3]} m_{\eta_8}^2
= \frac{2 [2]}{2 [2] - [3]} m_K^2 .                  \eqno(27)
$$
With appropriately fixed $q=q_{\rm ps}$ Eq.(27) is satisfied,
{\it without introducing singlet mixing angle},
if one puts the (mass of) physical $\eta$-meson in place of
$\eta_8$ - just this is meant in what follows.

Besides common feature of (26) and (27) (both give the
standard GMO in the corresponding limits
$\frac{F_K}{F_\pi} \to 1, \frac{F_\eta}{F_\pi} \to 1$
and $q \to 1$), there is essential difference:
the $q$-deformed one depends solely on $q$ whereas
Eq.(26) contains {\it two independent ratios}.
However, with an additional constraint
$$
1 + 3 F_\eta^2 / F_\pi^2 = 4 F_K^2 / F_\pi^2        \eqno(28)
$$
we have the juxtapositions
$$
\frac{F_K^2}{F_\pi^2} \leftarrow\!\rightarrow
\frac{\frac12 [2]}{2 [2] - [3]},
\hspace{20mm}
 \frac{F_\eta^2}{F_\pi^2}
\leftarrow\!\rightarrow
\frac{\frac13 [3]}{2 [2] - [3]}.                    \eqno(29)
$$
Hence, Eqs.(26) and (27) become correspondent to each other.

On the other hand, the ratio ${F_K}/{F_\pi}$
is related with the Cabibbo angle, see e.g., [24].
Due to (28), the same is true for $F_\eta / F_\pi$.
From this fact combined with the above correspondence (29) we
conclude: the realistic value $q_{\rm ps}$ is
{\it directly connectible} with the Cabibbo angle.

\smallskip
Now let us return to our $q$-dependent mass formulas for baryons:
Eq.(19) in the octet $\frac12^+$ sector and Eq.(21) in the decuplet
$\frac32^+$ sector. In our opinion, it is natural to extend to
these baryonic cases the conclusion just made
about possible connection $q\leftarrow\!\rightarrow \t_{\rm C}$.
In other words, we consider the values
$\t =\t_{\bf 10}\simeq\pi /14$ (decuplet case) and
$\t =\t_{\bf 8}=\pi /7$ (octet case) for deformation parameter
$q=e^{{\rm i}\t}$ as functions of $\t_{\rm C}$:
$\ \ \t_{\bf 10}=f(\t_{\rm C}),\ \ $  $\t_{\bf 8}=g(\t_{\rm C}).$

It is really surprising that \underline {the simplest choice}
$$  \t_{\bf 10}=\t_{\rm C}      \hskip 14mm
{\rm and} \hskip 14mm  \t_{\bf 8}=2\, \t_{\rm C}
$$
\vspace{-1mm}
provides excellent validity of mass sum rules (21) and (19).
Now, adopting $\t_{\bf 8}=2\, \t_{\bf 10}$ as {\it exact equality}
we get two implications:

(i) Since $\pi /7$ is strictly fixed value for $\t_{\bf 8},$ it is
tempting to suggest the exact value $\pi /14$ for the Cabibbo angle;

(ii) Excluding (due to equality $\t_{\bf 8}=2\, \t_{\bf 10}$)
the deformation parameter from the Eqs.(21) and (19) we obtain
the following new octet--decuplet mass formula:
\vspace{-1mm}
$$
\frac{m_{\O} - m_{{\Xi}^* } + m_{{\S}^* } - m_{\D} }
     { m_{\Xi^* } - m_{\S^* } }
= {\biggl ( 3 + \frac{ m_\Xi - m_\L }{ m_{\S }
              - m_N } \biggr ) }^{1/2},                \eqno(30)
$$
which is satisfied with remarkable precision.

\bigskip
\noindent {\bf 7. Concluding remarks}
\bigskip

\ni Let us recall again most important features of
the presented $q$-deformed hadron mass formulas:

(i) universality of the $q$-deformed decuplet mass formula, Eq. (21);

(ii) possibility of optimal choice (at strictly fixed
$q=e^{{\rm i}\pi /7},$ cf. Eq.(19)) from infinite set of
mass sum rules (18), thanks to degeneracy lifting caused by
the $q$-{\it deformation} of $SU(n_f)$;

(iii) topological meaning behind meson $1^-$ $q$-deformed
mass relations (3), embodied in knot invariants ascribed to
quarkonia and in possibility
to label flavors by a winding number.

(iv) in all three cases, $q$ is pure phase
(roots of unity); cases with baryons suggest a relation
$q\leftrightarrow \t_{\rm C}$ realized in simplest form, and
the exact value $\pi /14$ for $\t_{\rm C}$.

\smallskip
As shown in [25], it is possible to relate $\t_{\rm C}$ with ratio
of {\em subquark} (not quark) masses. However, the problem of
determination of (the value and genuine origin)
of $\t_{\rm C}$ {\it independently of values of quark masses} still
remains to be solved. Just in this context the idea that
space-time symmetries and/or internal symmetries may actually
appear through quantum groups/algebras, we hope, will be very
useful.

The final remark concerns appearance of knot invariants
in connection with vector mesons. The fact that knot
invariants are closely connected with quantum algebras
is well-known [26]. However, our use of $q$-algebras to the issue
of hadron masses and mass formulas gives a hint of which
\underline {concrete torus knots} are to be assigned to which
\underline {concrete} vector quarkonia.

\medskip
{\bf Acknowledgements.} The author is thankful to Prof. R.Jackiw
for valuable remarks and to Prof. H.Terazawa for useful discussions.
                     The research described in this
publication was made possible in part by the Award
No. UP1-309 of the U.S. Civilian Research \& Development
Foundation for the Independent States of the Former Soviet
Union (CRDF).

\end{document}